\documentclass[12pt,preprint]{aastex}
\singlespace
\def\p/{\mbox{$^1$}}
\def\pp/{\mbox{$^2$}}
\def\ppp/{\mbox{$^3$}}
\def\pppp/{\mbox{$^4$}}
\def\m/{\mbox{$^{-1}$}}
\def\mm/{\mbox{$^{-2}$}}
\def\mmm/{\mbox{$^{-3}$}}
\def\mmmm/{\mbox{$^{-4}$}}
\def\Ms/{\mbox{M$_\odot$}}

\slugcomment{Accepted for publication in Astrophysical Journal Letters}
\begin{document}

\title{The origin of the Local Bubble\altaffilmark{1}}
\shorttitle{The origin of the Local Bubble}

\author{Jes\'us Ma\'{\i}z-Apell\'aniz}
\affil{Space Telescope Science Institute\altaffilmark{2}, 3700 San Martin 
Drive, Baltimore, MD 21218, U.S.A.}

\altaffiltext{1}{Based on data from the {\em Hipparcos} astrometry
satellite.}
\altaffiltext{2}{The Space Telescope Science Institute is operated by the
Association of Universities for Research in Astronomy, Inc. under NASA
contract No. NAS5-26555.}

\begin{abstract}

The Sun is located in a low-density region of the interstellar medium 
partially filled with hot gas that is the likely result of several nearby 
supernova explosions within the last 10 Myr. Here we use astrometric 
data to show that part of the Scorpius-Centaurus OB association was 
located closer to the present position of the Sun $5-7$ Myr ago than today. 
Evolutionary synthesis models indicate that the association must have 
experienced $\sim$ 20 supernova explosions in the last $10-12$ Myr, a prediction
that is supported by the detection of four or five runaway stars escaping 
from it. The $\sim 6$ SNe produced by the Lower Centaurus Crux subgroup are 
likely responsible for the creation of the Local Bubble.

\end{abstract}


\section{Introduction}

The low-density region of the local ISM where the Sun 
is located is called the local cavity and is partially filled with hot 
($\sim 10^6$ K) low number density ($\sim 0.005$ cm\mmm/) coronal gas detectable
in soft X-rays \citep{Sfeietal99,Snowetal98}. The hot component of the local 
cavity is the Local Bubble (or LB) and is likely to have been produced by a 
series of several supernova (SN) explosions within the last 10 million years
(\citealt{SmitCox01} and references therein). 
Since the number density of early-type stars in the 
immediate solar neighborhood is very small \citep{Maiz01}, it appears unlikely 
that several nearby isolated massive stars would have exploded within such a 
short time period \citep{SmitCox01}. Furthermore, no OB associations of 
the right age exist within 100 pc of the Sun at this time, so the identity of 
the SN progenitors which produced the LB has remained hidden until now. The 
search for the culprits was severely hampered until recently by the 
impossibility of accurately tracing the past positions of the stars in the 
solar neighborhood, but the advent of the data produced by the ESA 
Hipparcos astrometry satellite \citep{ESA97} has changed the situation.

The local cavity has an approximate linear size of 200 pc and is slightly more 
elongated along the axis parallel to the Galactic rotation than along the 
radial axis, with fingers extending towards Galactic longitude $l = 310\degr$ 
and maybe also towards $l = 240\degr$. It is also more extended towards the 
North Galactic Pole (or NGP) than towards the South Galactic Pole (or SGP), 
being probably open-ended in the first direction \citep{Sfeietal99}. It has 
been mapped using different techniques, such as Na\,{\sc i} and H\,{\sc i} 
absorption \citep{Sfeietal99,Vergetal01}, dust extinction 
and polarization \citep{Fran90,Lero93}, and the distribution of EUV sources
\citep{Welsetal94}. The extension of the LB itself is more difficult to measure 
since it is only easily detected in soft X-rays \citep{Snowetal98,Sandetal98} 
and EUV broadband data \citep{Lieuetal93}, where confusion with other diffuse 
sources complicates the interpretation of measurements, and in O\,{\sc vi} 
absorption, where data towards only a few directions are available 
\citep{ShelCox94}. The non-detection of emission lines in the EUV 
\citep{Jelietal95,VallSlav98} suggests that a depleted or non-equilibrium 
$10^6$ K plasma is the dominant component of the LB \citep{SmitCox01}. The 
local cavity is surrounded by denser gas that is moving slowly 
\citep{Magnetal96} and, therefore, can be used to a first-order approximation 
as a fixed reference system. 

\section{Present and past positions of the Sco-Cen OB association}

The local cavity is not only poor in gas but also in hot stars. Only 
three O-B5 stars are found within 67 pc of the Sun and none of them is earlier 
than B2 \citep{Maiz01}. Recently, \citet{deZeetal99} have analyzed the census 
of the OB associations within 650 pc of the Sun position. The 
Scorpius-Centaurus OB association (or Sco-Cen) is the nearest one and can be 
divided into three subgroups \citep{Blaa64}: Lower Centaurus Crux (LCC), Upper 
Centaurus Lupus (UCL), and Upper Scorpius (or US). In Table~\ref{tab1} we show 
their ages as measured by \citet{deGeetal89}. In Fig.~\ref{fig1}(a) we show the 
present position of the center of each subgroup \citep{deZeetal99} and in 
Fig.~\ref{fig1}(b) we show a blowup with the position of the OB stars with 
well-determined distances in each subgroup using the procedure established by
\citet{Maiz01}. In order to analyze whether some of the former 
members of Sco-Cen which must have already exploded as SN may have 
contributed to the creation of the LB, we used the membership lists and the 
data presented by \citet{deZeetal99} to calculate the positions of the center of
each subgroup in the past. We used a coordinate system which is centered at the 
present Sun position, moves with the Galaxy at its local rotation speed (i.e. 
it is a local standard of rest), and has $x$, $y$, and $z$ defined by the 
direction of Galactic rotation, the outer radial direction, and the NGP, 
respectively. Velocities of the center of the subgroups have to be corrected 
for the effects of solar motion \citep{DehnBinn98} and Galactic rotation 
\citep{FeasWhit97}. The motion of a star in the $z$ direction can be described 
as a harmonic oscillation with a period of 83 Myr and centered at the Galactic 
plane \citep{King96,HolmFlyn00}. The motion in the $x$-$y$ plane has a more 
complicated trajectory but it can be approximately described by a retrograde 
ellipse with a period of 167 Myr and an axis ratio of 1.48, so that the motion 
of a star from a non-rotating point of view high above the Galaxy resembles an 
epicycle \citep{King96,FeasWhit97}. Taking into account those effects, we have 
computed the positions of the three subgroups 5 Myr ago and the positions of 
UCL and LCC 10 Myr ago (US did not exist at that time). The results are shown 
in Table~\ref{tab2} and plotted in Fig.~\ref{fig1}(a). We also show in 
Fig.~\ref{fig1} the position of the Ophiucus molecular cloud, as deduced from 
CO and IRAS data \citep{deGeetal89,Lore89a,Lore89b}. 

\section{Number of supernova explosions}

It is clear from Fig.~\ref{fig1}(a) that the Sco-Cen OB association (especially 
the LCC subgroup) was closer to the present Sun position $\sim 5$ Myr ago than 
today, which makes it a likely source for the $\gtrsim 3$ SN believed to be
needed to produce the LB. As a first step to test this hypothesis, we need to 
verify how many SN have been produced in Sco-Cen within the last 10 Myr. In 
order to do that, we counted the number of present O-B2.5 and B3-B9.5 in each 
subgroup using the membership lists of \citet{deZeetal99} and the spectral 
classifications of the Hipparcos \citep{ESA97} and Michigan \citep{HoukSwif99} 
catalogs (Table~\ref{tab1}). Main sequence O-B2.5 stars are the ones which 
end up their lives exploding as SNe, since a B2.5 V star has a mass of 
$\approx 9$ \Ms/. We compared the observed numbers of early/late OB stars with 
those predicted by Starburst 99 models \citep{Leitetal99} for stellar groups of 
the ages given in Table~\ref{tab1} with solar metallicity, Salpeter IMF, and an 
upper mass limit of 100 \Ms/, and obtained the number of expected previous SNe. 
The first SN is expected to take place when the stellar group is 3-5 Myr 
old\footnote{The exact moment depends on the mass of the most massive star,
which for relatively low-mass subgroups such as those in Sco-Cen is quite 
uncertain due to the stochastic nature of the star formation process.}
and then the rest of the SNe take place at an approximately constant rate 
during the next $\sim 30$ Myr. The results are also shown in Table~\ref{tab1}.

	The proportions of early-to-late OB stars in LCC and US are well 
fitted by the Starburst 99 predictions, so the expected number of SNe for those 
subgroups is probably quite reliable. The predictions are also in agreement 
with our knowledge of the US subgroup, where one SN took place 1 Myr ago 
\citep{Hoogetal01}, giving birth to the pulsar PSR J1932+1059 and ejecting the 
runaway star $\zeta$ Oph, and a red supergiant, Antares, is due to explode 
within the next few hundred thousand years. The proportion of OB stars in UCL 
is less well fitted by the model, probably indicating an anomalous IMF at upper 
masses (a phenomenon which is not at all surprising, since the small expected 
number of O-B2.5 stars can easily lead to stochastic fluctuations). However, 
even if we take the conservative approach of considering only half of the 
predicted SNe in UCL, there should still be $\sim 6$ SNe there exploding 
within the last $10-12$ Myr plus another $\sim 6$ SNe in LCC in the last $7-9$ 
Myr. These minimum values are supported by the detection of three or four 
additional runaway B1-B5 stars that appear to have been ejected (two from LCC 
and one or two from UCL) as a result of the explosion of their companions as 
SNe \citep{Hoogetal01}. The LCC runaways were ejected 2.5 Myr and 4.0 Myr ago,
the certain UCL runaway 8.0 Myr ago, and the possible UCL runaway 3.0 Myr ago.

\section{The history of the Local Bubble}

	Since there are so many predicted Sco-Cen SNe in the last $10-12$ Myr,
the question maybe should be not whether that association is responsible for the
LB but rather why is not the bubble larger and approximately centered at its 
position of $\sim 5$ Myr ago (i.e. its weighted mean position within its 
SN-producing lifetime). A likely explanation is provided by
\citet{Fris98}: The Ophiucus molecular cloud (the remnant of the progenitor of
Sco-Cen) and the larger Aquila Rift are located towards the first Galactic
quadrant and would have impeded the expansion of the bubble towards $+x$ and
$-y$ (at least close to the Galactic plane). On the other hand, the expansion 
towards $+y$ would have been facilitated by the existence of an interarm region
there. We should also consider that the Ophiucus molecular cloud probably 
shares its general motion with Sco-Cen (this cannot be proven with the data 
currently available since its proper motion is unknown but its radial velocity 
\citep{Lore89b} is consistent with this statement). We would then expect 
that the motion of the molecular cloud in the negative $x$ direction would have 
made it occupy now part of the space previously held by the coronal gas, thus
helping to create the partial Na\,{\sc i} ridge between the LB and the Loop I 
superbubble, which appears in Fig.~\ref{fig1} as the minimum around 
$(-75,-150)$; note that the density contours there are probably not very
precise.

	The picture that emerges from the data in this letter can then be
summarized as follows. The LCC and UCL subgroups have produced enough SNe to
create both the Local Bubble and the Loop I superbubble. Given their past
positions, the two bubbles likely started as a single entity, with the SNe in 
LCC being primarily responsible for the expansion towards $+y$ (the current LB)
and the SNe in UCL for the expansion towards $-x$ (the current Loop I
superbubble). The $\sim 6$ LCC SNe would have exploded within a 
$\sim 40$ pc-wide band of the trajectory of its center shown in 
Fig.~\ref{fig1}(a), thus creating the LB part of the original Local Bubble-Loop
I superbubble. Since LCC is currently abandoning the local cavity and any recent
SN would have exploded quite far from its center, we would expect the dense 
material around it to have started to fill it up again \citep{SmitCox01}. That 
is what Fig.~\ref{fig1} suggests, since the current extent of the LB just grazes
the past trajectory of LCC\footnote{Note, however, that 1 or 2 of the LCC SNe
could have actually exploded within the present extent of the LB due to the
finite extent of the subgroup.}. If the LB appears to be in the last stages of
its evolution, the forecast for the Loop I superbubble is more optimistic. Not 
only is most of UCL still inside it but LCC is moving towards it and so is the 
younger subgroup US. All together, they are expected to produce $\sim 35$ SNe 
within the next $25-30$ Myr, so its short-term survival seems to be more
likely than that of the LB. Indeed, a recent SN explosion took place there and 
its remnant is expanding towards us at this time 
\citep{EggeAsch95}\footnote{Those authors propose that the partial Na\,{\sc i} 
ridge between the LB and the Loop I superbubble is the result of the interaction
between the two of them produced by the most recent SNe.}.

	Therefore, we can conclude that LCC was at the 
right place, at the right time, and experienced enough SNe to produce the LB.
No other nearby OB association can be traced back in time to 
be in such a situation according to the data presented by \citet{deZeetal99}. 
Other authors \citep{Fris81,Breietal96} have considered the possibility that 
the LB was created by Sco-Cen as a bubble expanding from its present 
position (i.e. as a blister of the Loop I superbubble). 
However, we have seen here that there is no need for such an explanation: 
most of the LCC SNe exploded much closer to 
the present position of the Sun and the original bubble may have been larger 
than its present size since the motion of the Ophiucus molecular cloud is 
probably closing the connection between the LB and the Loop I superbubble.
The largest uncertainties in this picture lie in the extent of the
original molecular cloud and in its current 3-D motion; the latter question 
could be answered when ALMA \citep{Brow00} is built. 
Finally, we also have to mention that
it cannot be excluded that a single (or even two) SN from the diffuse disk
population may have contributed to the formation of the LB. However, as 
estimated by \citet{SmitCox01}, the chances of having several of them exploding 
in our vicinity within a few Myr are rather low.

\section{Implications}

Apart from the relevance of the present study to the knowledge of the solar
neighborhood, we would like to point out its implications with respect to the 
overall porosity of the ISM. Two competing theories suggest that the general 
morphology of the ISM is that of a ``bubble bath'' (\citealt{McKeOstr77}, with 
hot bubbles occupying most of the available volume) or that of a ``Swiss 
cheese'' (\citealt{SlavCox93}, with hot bubbles occupying a smaller fraction of 
the volume). The fact that our Local Bubble can be traced back to a known OB 
association indirectly supports the ``Swiss cheese'' model, since it shows that 
$\sim 6$ SNe were unable to create a much larger bubble.  

\begin{figure}
\plotone{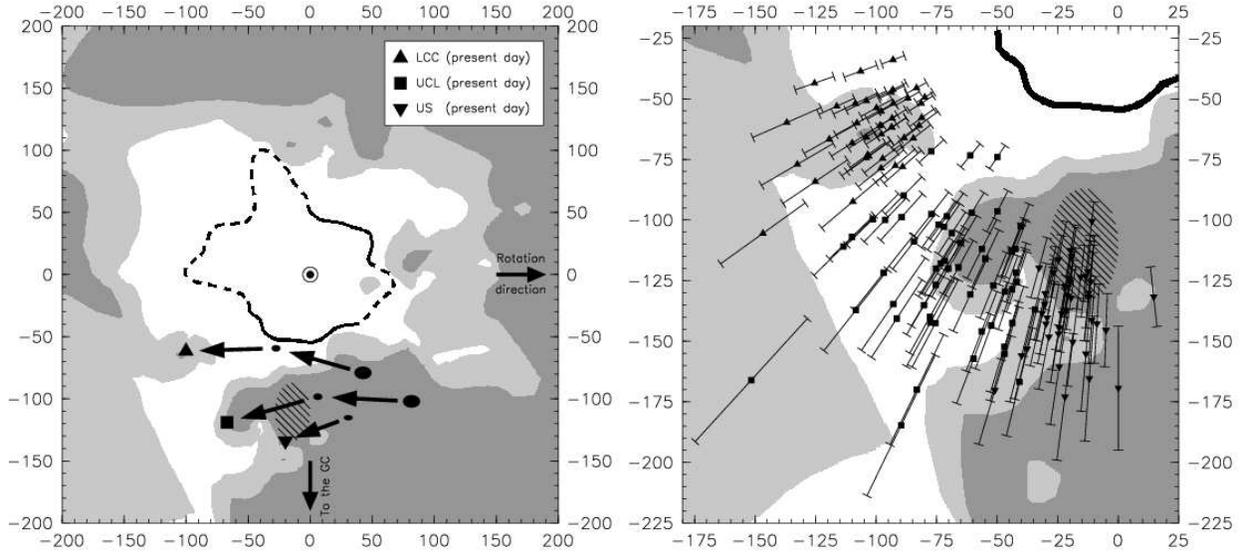}
\caption{(a) The local cavity and LB in the plane of the Galactic Equator. The 
filled contours show the Na\,{\sc i} distribution \citep{Sfeietal99}, with 
white used for low-density regions and dark grey for high density ones. The 
black contour shows the present size of the LB as determined from X-ray data 
\citep{Snowetal98}, with the dashed lines indicating contaminated areas where 
the limits of the LB cannot be accurately determined. The diagonal-line-filled 
ellipse shows the approximate position of the Ophiucus molecular cloud 
\citep{deGeetal89,Lore89a,Lore89b}. The present and past $x$ and $y$ 
coordinates of the center of the three subgroups of the Sco-Cen association are 
shown. For LCC and UCL the past positions shown are those of 5 Myr and 10 Myr 
ago while for US only the position of 5 Myr ago is shown. The dimensions of 
the solid-filled ellipses indicate the uncertainties in the past positions. 
Coordinates are expressed in pc. (b) Blowup of the left figure with the present 
positions of the OB stars in each of the three subgroups. Only those stars with 
accurately determined positions are shown. The symbol used in each case
indicates the subgroup membership using the code established in the left panel.}
\label{fig1}
\end{figure}

Another consequence of this study is that the proximity of the Sco-Cen 
association has increased by a significant factor the rate of SN explosions 
within 150 pc of the Sun in the last 6 Myr with respect to the mean rate in the
last $\sim 100$ million years\footnote{Note, however, that the increase is not 
as large as what could be deduced from a first look at Fig.~\ref{fig1}(a) 
because the Sun is moving from the upper left quadrant there.}. In a follow-up 
paper \citep{Benietal01} we explore the possible geological and biological 
consequences of those nearby SNe.

\acknowledgments

	The author would like to thank Santiago Arribas and an anonymous referee
for useful comments on improving this paper. Support for this work was provided 
by grant 82280 from the STScI DDRF. 

\bibliographystyle{apj}
\bibliography{ms}

\begin{deluxetable}{lccccccc}
\tablecaption{Observed and expected number of OB stars and SNe in each 
subgroup. The ages were obtained from \citet{deGeetal89} and the measured
numbers from \citet{deZeetal99} and \citet{Hoogetal01}. The ranges given for
the measured star numbers arise from uncertainties in the membership lists and 
spectral classification while the measured numbers of SNe are the lower limits
given by the known runaway stars. The expected numbers were obtained using 
Starburst 99 models \citep{Leitetal99} with solar metallicity, Salpeter IMF, and
an upper mass limit of 100 \Ms/. The expected number of previous SNe was 
calculated from the total number of OB stars and the age.\label{tab1}}
\tablewidth{0pt}
\tablehead{\colhead{Subgroup} & \colhead{Age} & 
 \multicolumn{3}{c}{Measured number} & \multicolumn{3}{c}{Expected number} \\
 & \colhead{(Myr)} & \colhead{O-B2.5} & \colhead{B3-B9.5} & \colhead{SNe} & 
 \colhead{O-B2.5} & \colhead{B3-B9.5} & \colhead{SNe}} 
\startdata
LCC & $11-12$ &  $5-7$  & $31-35$ & $\ge 2$ &  7 & 32 &  6 \\
UCL & $14-15$ & $14-17$ & $44-49$ & $\ge 2$ & 10 & 52 & 13 \\
US  &  $5-6$  & $12-15$ & $30-36$ & $\ge 1$ & 11 & 36 &  1 \\
\enddata
\end{deluxetable}

\begin{deluxetable}{lcrrr}
\tablecaption{Present and past positions of the three subgroups in Sco-Cen. The 
positions refer to the center of each subgroup and use the coordinate system 
defined in the text. No position is given for US 10 Myr ago because the 
subgroup did not exist at that time.\label{tab2}}
\tablewidth{0pt}
\tablehead{\colhead{Subgroup} & \colhead{$t$}       &  \colhead{$x$}  &  
           \colhead{$y$}      & \colhead{$z$}       \\
                              & \colhead{(Myr ago)} &  \colhead{(pc)} &  
           \colhead{(pc)}     & \colhead{(pc)}}
\startdata
LCC &  0 & $-100$ &  $-62$ & $10$ \\
    &  5 &  $-28$ &  $-59$ &  $3$ \\
    & 10 &   $43$ &  $-79$ & $-8$ \\
UCL &  0 &  $-67$ & $-119$ & $31$ \\
    &  5 &    $6$ &  $-98$ & $15$ \\
    & 10 &   $82$ & $-102$ & $-7$ \\
US  &  0 &  $-20$ & $-134$ & $52$ \\
    &  5 &   $31$ & $-115$ & $45$ \\
\enddata
\end{deluxetable}

\end{document}